\documentclass[
aip,
amssymb,
amsmath,
reprint,
]{revtex4-1}
\usepackage{everypage}
\usepackage{amsmath}
\usepackage{amssymb}
\usepackage{nicefrac}
\usepackage%
{graphicx}
\usepackage{extdash}
\usepackage{epsfig}
\usepackage{textcomp}
\usepackage{letltxmacro}
\makeatletter
\let\oldr@@t\r@@t
\def\r@@t#1#2{%
\setbox0=\hbox{$\oldr@@t#1{#2\,}$}\dimen0=\ht0
\advance\dimen0-0.2\ht0
\setbox2=\hbox{\vrule height\ht0 depth -\dimen0}%
{\box0\lower0.4pt\box2}}
\LetLtxMacro{\oldsqrt}{\sqrt}
\renewcommand*{\sqrt}[2][\ ]{\oldsqrt[#1]{#2} }
\makeatother
\DeclareFontFamily{U}{euc}{}
\DeclareFontShape{U}{euc}{m}{n}{<-6>eurm5<6-8>eurm7<8->eurm10}{}%
\DeclareSymbolFont{AMSc}{U}{euc}{m}{n} 
\DeclareMathSymbol{\umu}{\mathord}{AMSc}{"16} 

\renewcommand{\vec}[1]{\boldsymbol{#1}}
\newcommand{\ensuretext}[1]{\ensuremath{\text{#1}}}

\newcommand{\me}{\ensuremath{m_{\mathrm{e}}}}

\newcommand{\unit}[1]{\ensuretext{\textrm{\,}}\ensuremath{\mathrm{#1}}}
\newcommand{\eV}{\textit{e}\mathrm{V}}
\newcommand{\MeV}{\mathrm{M}\eV}
\newcommand{\keV}{\mathrm{k}\eV}

\newcommand{\Mum}{\ensuremath{\umu}\ensuremath{\mathrm{m}}}
\newcommand{\mum}{\textrm{\,\ensuremath{\mathrm{\Mum}}}}

\begin{document}

\title{Simple scaling equations for electron spectra, currents and bulk heating in ultra-intense short-pulse laser-solid interaction}

\author{T.~Kluge}
\email[]{t.kluge@hzdr.de}
\homepage[]{http://hzdr.de/crp}
 \affiliation{Helmholtz-Zentrum Dresden-Rossendorf, 01328 Dresden, Germany}

\author{M.~Bussmann}
 \affiliation{Helmholtz-Zentrum Dresden-Rossendorf, 01328 Dresden, Germany}


\author{Ulrich Schramm}%
\author{Thomas E. Cowan}%
 \affiliation{Helmholtz-Zentrum Dresden-Rossendorf, 01328 Dresden, Germany}
 \affiliation{Technische Universit\"at Dresden, 01069 Dresden, Germany}%

\date{\today}

\begin{abstract}
Intense and energetic electron currents can be generated by ultra-intense lasers interacting with solid density targets. 
Especially for ultra-short laser pulses their temporal evolution needs to be taken into account for many non-linear processes as instantaneous values may differ significantly from the average. 
Hence, a dynamic model including the temporal variation of the electron currents -- which goes beyond a simple bunching with twice the laser frequency but otherwise constant current -- is needed. 
Here we present a time-dependent solution to describe the laser generated currents and obtain simple expressions for the electron spectrum, temporal evolution and resulting correction of average values. 
To exemplify the semi-empiric model and its predictive capabilities we show the impact of temporal evolution, spectral distribution and spatial modulations on Ohmic heating of the bulk target material. 
\end{abstract}

\pacs{12345}

\maketitle
\section{Introduction}
The generation of hot and dense electron currents by ultra-intense laser irradiation of solids has important impact on the heating of the solids~\cite{Sentoku2007,*green2007surface,*Nishimura2011,*LGHuang}, generation of resistive fields~\cite{d2009importance,*Sentokuresistive,*Chawla2013,Leblanc2014} development of current instabilities~\cite{Califano2002,*Fuchs2003,*Wei2004a,*Manclossi2006,*Quinn2012} and eventually the acceleration of ions at the plasma rear surface~\cite{PhysRevLett.84.4108,*Snavely-IntenseProtonBeams}. 
Its exact time-dependent calculation is of high relevance, e.g. due to non-linearities in the respective equations. 
Since there exists no complete and self-consistent explicit theory for the current generation from ultra-intense laser interaction with solids, it is often convenient to use heuristic arguments and general conservation laws. 
It is very common to calculate the laser generated current of energetic electrons $j_h(t)=n_h(t) v_h(t)$ employing the energy density flux conservation equation. 
Surprisingly, in literature the exact time-dependent form is only rarely used. 
It is more customary to use the much simpler form~\cite{haines, Leblanc2014}
\begin{equation}
\chi \frac{a_0^2}{2} = \left(\left\langle\gamma_h(t)\right\rangle-1\right) \left\langle n_h(t) \right\rangle \left\langle v_h(t)\right\rangle,
\label{eqn:EfluxconserveSimpler}
\end{equation}
setting the cycle averaged absorbed laser intensity equal to the product of the average energy, density and velocity of accelerated electrons. 
Here, $\left\langle...\right\rangle$ denotes the time average over a laser period, $a_0=\sqrt{2I/n_c\me c^3}$ (where $I$ is the laser intensity, $n_c=\me\epsilon_0\omega_0^2e^{-2}$ is the critical density with $\omega_0$ being the laser frequency, $m_e$ the electron rest mass, $e$ the electron charge, and $\epsilon_0$ the electric constant), $\chi a_0^2/2$ is the energy flux absorbed into energetic electrons (which may be differing significantly from total laser absorption) and $\gamma_h(t)$, $n_h(t)$, $v_h(t)$ are the time dependent energy, density and velocity in laser direction of laser accelerated electrons. 
Here and in the following we use dimensionless units, $c=\me=n_c=\omega_0=e=1$. 
Hence, velocities, densities, energies and currents are given in units of $c$, $n_c$ $(1.1\cdot 10^{21}\lambda^{-2} \unit{cm}^{-3})$, $\me c^2$ $(511\unit{\keV})$ and $e n_c c$ $(48\unit{kA/\mum}^{2})$, respectively. 
Also, since throughout the present work we only consider a transversely homogeneous energy flow along one direction, all values we refer to can be and are meant as values normalized to the unit area, i.e. areal densities taken across the 2 transverse directions. \\
With the above equation, setting the laser absorption coefficient $\chi\equiv 1$ and assuming a ponderomotive-like scaling for the average hot electron energy\cite{Wilks}, 
\begin{equation}
\langle\gamma_h\rangle=\sqrt{1+\frac{a_0^2}2}
\label{eqn:ponderomotive}
\end{equation}
one quickly finds 
\begin{equation}
\langle j_h\rangle\cong \langle\gamma_h\rangle n_c c
	\label{eqn:j_h}
\end{equation}
for $a_0 \gg 1$ which coincides with heuristic arguments~\cite{Fuchs2003}. 
This equation holds also for $\chi\le 1$ with an \textit{ad hoc} modified hot electron scaling, replacing $\langle\gamma_h\rangle$ with $\bar\gamma_h=\sqrt{1+\chi\frac{a_0^2}2}$\cite{Leblanc2014}. 

Obviously, equation~\eqref{eqn:EfluxconserveSimpler} is not correct if instantaneous values differ significantly from their average values. 
However, this is expected for ultra-intense laser interaction with plasmas especially at  ultrashort pulse duration (few tens of femtoseconds)~\cite{popescu:063106}. 
Contrarily, for sufficiently long laser pulses the phase space of electrons may become mixed during the irradiation, e.g. due to electron recirculation\cite{Mackinnon2002,Kluge2010}, so that the instantaneous and local energies and velocities of electrons are not a specific single value anymore but resemble a broad distribution (i.e. the phase space volume is large, not peaked) that does not vary in time much anymore. \\
In this paper we limit ourself to the situation of an undisturbed initially cold plasma that is excited only by a few isolated laser periods. 
Then, as will be shown, the phase space distribution is very narrow and varies strongly during a laser period. 
E.g., considering the simple example of electrons bound at the front surface of a plasma slab before being pushed into the plasma one may expect an oscillatory temporal evolution of the electron motion. 
In such a case the temporal evolution especially of $\gamma_h(t)$ and $n_h(t)$ need to be considered and the averages be taken correctly, 
\begin{equation}
\chi \frac{a_0^2}{2} = \left\langle\left(\gamma_h(t)-1\right) n_h(t) v_h(t)\right\rangle
\label{eqn:Efluxconserve}
\end{equation}
In the following we outline the procedure to derive the laser generated electron current $\left\langle j_h\right\rangle$ and $j_h(t)$ and spectrum from Eq.~\eqref{eqn:Efluxconserve} without such \textit{ad hoc} assumptions and give specific expressions for important exemplary cases. 
We compare the results to particle-in-cell (PIC) simulations and demonstrate the relevance for laser plasma heating and resistive magnetic field generation. 
It will be shown that not only the temporal evolution matters, but also spectral shape of the electron bunches as well as spatial structure (e.g. filaments) and dispersion. \\

\section{Simulations}
We performed 2 dimensional collisional PIC simulations using the code PICLS\cite{picls} to obtain a benchmark for the analytic model for $j_h(t)$. 
In order to be as close to the specific model requirements as possible we employ a spatially plane laser pulse with a very short rise time of only $T/2$ followed by a flat top, spatially constant in the directions transverse to the propagation direction. 
The plasma slab was modeled by a flat foil with the thickness of 2 laser wavelengths $\lambda_0$ and consisted of a preionized neutral ion-electron mixture with a density of $400 n_c$ each. 
The ion charge-to-mass ratio was set to $Q/A=1/2$ (with $A=1836\,\me$) and the slab was covered with a small exponential preplasma with scale length $0.1\lambda_0$ at the front surface. 
For the following analysis we consider only those electrons originating from the direct laser interaction, i.e electrons originating from the preplasma or skin-depth layer at the plasma front surface and average over a region of interest (ROI) comprising a distance of $1\lambda_0$ in laser direction starting $0.5\lambda_0$ inside the plasma and extending over the full simulated volume in transverse direction, hence we may expect to comprise two electron bunches from the $2\omega_0$ laser Lorentz force. 
All data is taken at the time when the first bunch accelerated after the laser pulse envelope is at maximum at the plasma front surface reaches the rear of the ROI.\\ 
\begin{figure}
	\includegraphics[width=8.5cm]{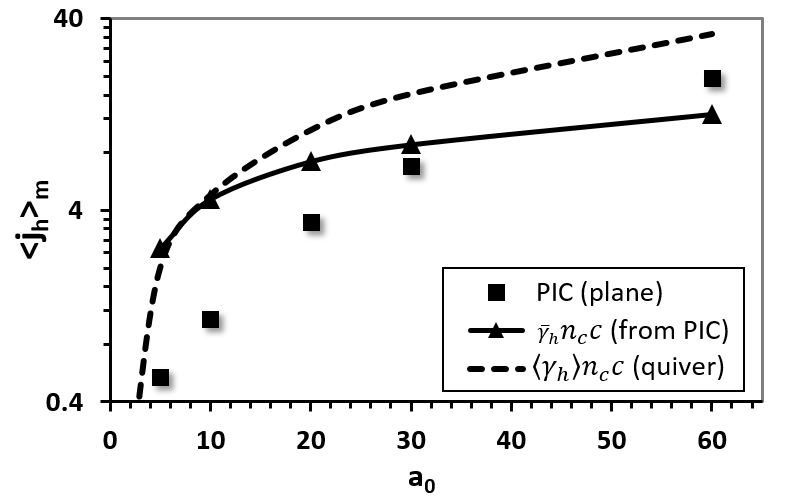}
	\caption{Current density averaged in transverse direction from PIC simulations with a plane laser wave front, compared to the popular scaling law $j_h=\gamma n_c c$ with $\gamma=\bar\gamma$ being the ensemble average electron energy from PIC (solid line) and $\gamma= \langle\gamma_h(t)\rangle$ with $\gamma_h(t)$ being the quiver energy Eqn.~\eqref{eqn:gammat} (dashed line).}
	\label{fig:j_total}
\end{figure}

Fig.~\ref{fig:j_total} shows the average current in laser direction of laser accelerated electrons inside the ROI compared to the analytical scaling Eq.~\eqref{eqn:j_h}. For the dashed line we used the average quiver energy of electrons in the transverse electric laser field (i.e. essentially the ponderomotive scaling), for the solid line we replaced $\left\langle\gamma_h\right\rangle$ with the average electron energy from the PIC simulation in the ROI, $\bar{\gamma}_h$. 
It is readily seen that the popular scaling~\eqref{eqn:j_h} does not describe the simulation results well. 
Using the quiver energy scaling, the current is consistently overestimated, with the average energy from the simulation Eqn.~\eqref{eqn:j_h} overestimates the current for $a_0\le 20$ while it significantly underestimates it for larger $a_0$. 
The differences are rather large, ranging up to an order of magnitude. 

It is important to note that here and in the following we only consider electrons with kinetic energy larger than $m=\gamma-1=1$, both for the simulation and for the analytic estimate. 
In the ROI there are not only electrons from two $2\omega_0$ bunches but due to dispersion also slow moving electrons from earlier bunches. 
By setting the energy threshold we can filter them out to a large extend.

\begin{figure}
	\includegraphics[width=8.5cm]{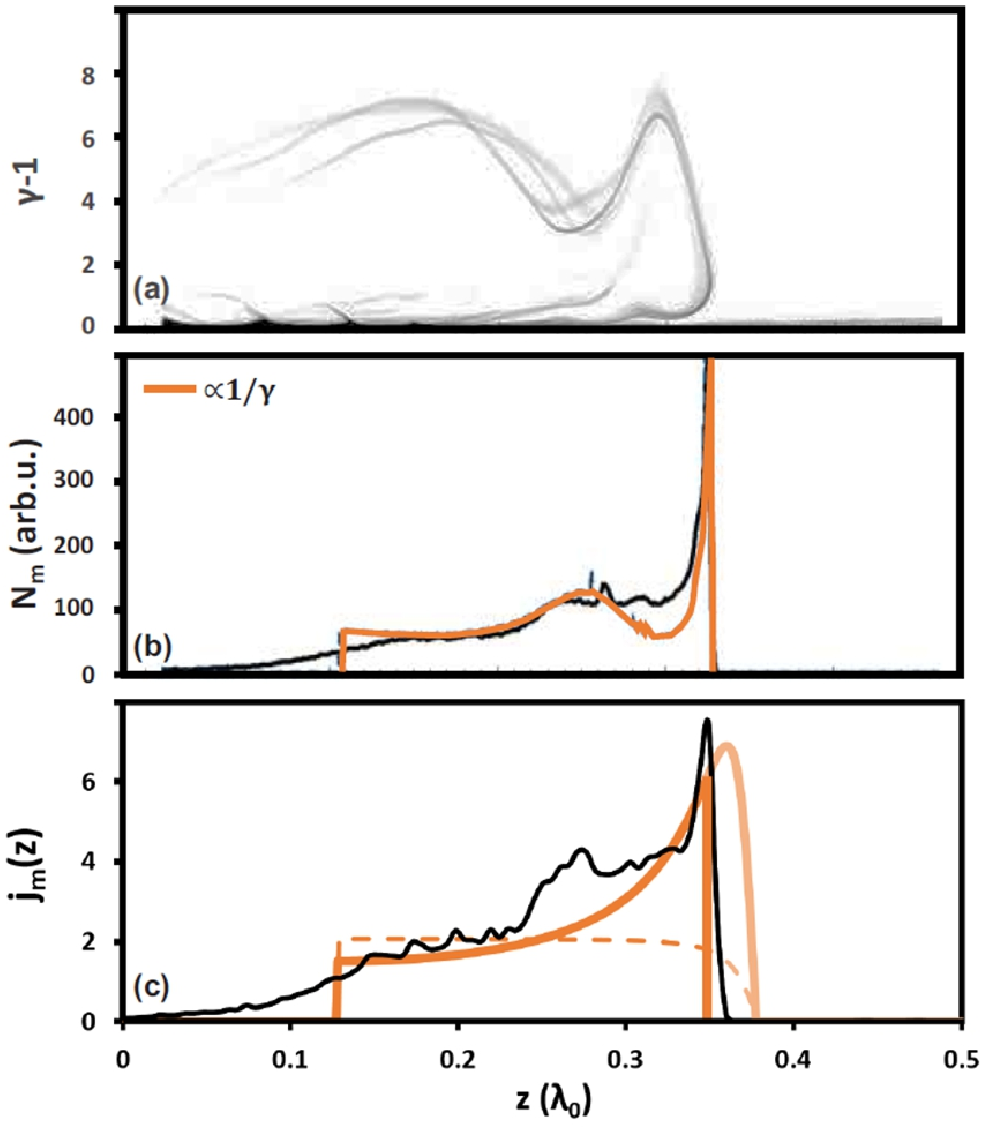}
	\caption{Spatial profile of energy (a), density (b) and current (c) of laser accelerated electrons along the laser axis (laser incident from negative direction, $a_0=10$, target starts at $z=0$). For (b) and (c) only electrons with $\gamma-1>m\equiv 1$ were considered. In (a), the first 4 bunches are shown on top of each other while (b) and (c) show the average over those 4 bunches. The orange line in (b) is a fit with $1/\gamma$ with $\gamma$ from (a). The orange lines in (c) are model predictions from Eqn.~\eqref{eqn:j_final_temp} (solid) and~\eqref{eqn:AvgCurrPond} (dashed).}
	\label{fig:j_summary}
\end{figure}

The reason why the analytical model Eq.~\eqref{eqn:j_h} fails in describing the PIC results of course is the strong variation of the electron energy, density and hence current with time. 
We showed and discussed the temporal evolution of the energy within one of the $2\omega_0$ bunches in~\cite{kluge2016controlled}, here we also show the evolution of the density and laser generated current in Fig.~\ref{fig:j_summary} exemplary for the simulation with $a_0=10$. 
The shape of all quantities were found to be very stable at least between the first four subsequent bunches, with only negligible cycle-to-cycle variation. 
For example the energy evolution follows the quiver energy evolution with the exception a dip and a peak at z=0.4 and 0.3 which was found to be due to interaction with the target bulk, i.e. a strong transient longitudinal field present at the front surface of the plasma at the time when the respective part of the bunch passes the critical density surface. 
As can be seen, in density and current this influence is much less pronounced, so we will neglect it in modeling in the following. 

\section{Spectrum of laser accelerated electrons}
\begin{figure*}
	\includegraphics[width=14cm]{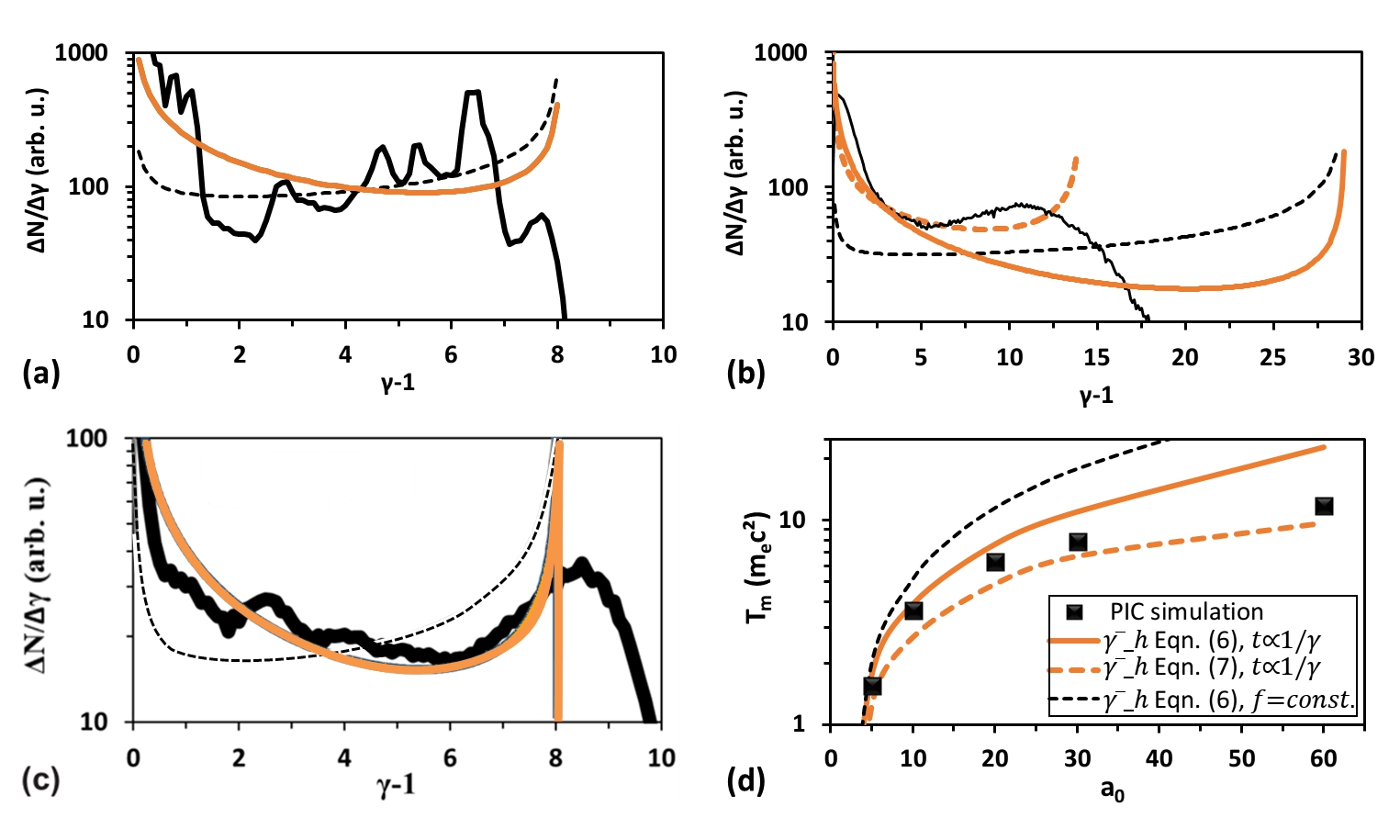}
	\caption{Electron spectra for all electrons in the first bunch in the ROI defined in the main text (a), (b) and in front of the target (c). The laser field vacuum amplitude is $a_0=10$ (a,c) and 30 (b). The spectrum~\eqref{eqn:spectrum} assuming $\hat p_{osc}=0.9 a_0$ (in agreement with simulations) is shown by the orange solid (black dashed) line for $f\propto1/\gamma$ ($f=const.$) and $\langle n_h\rangle$ extracted from the simulation. Orange dashed line in (b) shows spectrum~\eqref{eqn:spectrum_long} predicted for ultra-relativistic laser intensity. (d) Average energy of energetic laser accelerated electrons from simulations (black dots)  
compared to $T_{m}$ from Eqn.~\eqref{eqn:T_m}, using different expressions for $\gamma_h$ and $f$. A clear transition from the short scale regime (orange solid line) to the long scale regime (orange dashed line) can be seen. Quiver energy and $f=const.$ (black dashed line) does not fit the simulations. } 
	\label{fig:spec}
\end{figure*}

The spectrum of laser accelerated electrons injected at the front target surface can be simply  estimated when their energy and density are known as a function of time over a laser period. 
The spectrum then reads\begin{equation}
 \frac {dN_h}{d\gamma_h}=N_h^{tot}\frac{f(t)}{\dot{\gamma_h}}.
 \label{eqn:spectrum_general}
\end{equation}
For convenience, here we rewrite $\partial N_h/\partial t=N_h^{tot}f(t)$ where $N_h^{tot}$ is the total number  of electrons accelerated by the laser in one bunch of duration $\tau=\lambda/4=\pi/2$. 
Note that $f(t)$ is normalized to $\int_0^\tau f(t) dt = 1$.  
We are left with determining $\gamma_h(t)$, $f(t)$, and $N_h^{tot}$. 

The laser action on the target electrons can be divided into a transverse force by the electric field and a longitudinal force by the magnetic field. 
The former constitutes the driver for a transverse harmonic electron oscillation $p_{osc}=\hat{a}_0\sin(\omega_0 t)$ (where $\hat{a}_0$ is the laser electric field strength at the plasma front surface). 
The latter first pulls electrons moving in transverse direction off the target surface into the vacuum direction and upon reversal of the sign of the field pushes them back. 
For \textit{mildly relativistic} laser amplitudes (i.e. $a_0\lesssim 20$) the longitudinal amplitude is small and the amount of energy an electron can obtain while co-moving with the field is small also, the temporal energy evolution of an electron follows the quiver energy
\begin{equation}
 \gamma_h(t)=\sqrt{1+{p_{osc}}(t)^2},
 \label{eqn:gammat}
\end{equation}
i.e. the usual absorption-independent expression. 

For large laser strengths or a target with long preplasma (more than few hundred nanometers) we have to drop the restriction of considering electrons to only oscillate transversely. 
We have shown in~\cite{Kluge2011a} that then $\gamma_h$ can be obtained by
\begin{align}
\label{eqn:gammat_long}
\gamma_h(t)=&\nicefrac{S(t)^2}{2}+\nicefrac{2}{S(t)^2-1}\\
S(t)=&\sqrt[3]{\sqrt{9p^2_{osc}(t)+8}+3p_{osc}(t)}. 
\nonumber \end{align} 
In the following we will refer to this case as the \textit{ultra-relativistic} case. 

It is important to point out that often $\hat{a}_0\cong a_0$ is a good choice in experimentally relevant situations since for many high contrast short pulse laser systems the overcritical short preplasma scale length is around $0.1\lambda_0$. 
In this case the approximate equality between $a_0$ and $\hat a_0$ can be shown analytically from Maxwell equations for non-relativistic intensities~\cite{Roedel2012} and for $a_0>1$ it can be shown by simulations that this approximation remains valid over a wide range of experimentally relevant intensities. 
This means that in contrast to \cite{Leblanc2014} where $\hat a_0 \equiv \chi a_0^2$ was imposed we assume the laser field amplitude to be invariant of the laser absorption $\chi$; hence $\chi$ can only affect the number of accelerated electrons. \\

We will also discuss two expressions for $f(t)$: $f(t)=const.=\tau^{-1}$, and, again following~\cite{Kluge2011a}, 
\begin{equation}
f(t)=\frac{f_0}{\gamma_h(t)}
 \label{eqn:ft}
\end{equation}
with $f_0=\left(\tau \langle 1/\gamma_h\rangle\right)^{-1}$. 
With the quiver energy Eqn.~\eqref{eqn:gammat} for the former the ensemble average energy of the laser accelerated electrons would be
\begin{eqnarray}
\nonumber
\bar\gamma_h&=&\nicefrac{\langle \gamma_h(t) N_h(t)\rangle}{\langle N_h(t) \rangle} = \langle \gamma_h \rangle\\
&=&2E(i\hat a_0)/\pi
\label{eqn:avg_energy_pond}
\end{eqnarray}
(where $E(x)$ is the complete elliptic integral of the second kind), which approximately reproduces the Wilks ponderomotive scaling. 
For the latter the average energy is 
\begin{eqnarray}
\bar\gamma_h&=&\langle \nicefrac{1}{\gamma_h(t)}\rangle^{-1}\\
&=&\pi/2K(i\hat a_0)
\label{eqn:avg_energy_pre}
\end{eqnarray}
where $K(x)$ is the complete elliptical integral of the first kind, hence it is $f_0=\bar \gamma/\tau$. 

In Fig.~\ref{fig:j_summary}b the evolution of the number of laser accelerated electrons with an energy of $\gamma_h-1>m\equiv 1$ is shown for the PIC simulation with $a_0=10$ together with a fit of $1/\gamma_h$ from the same simulation (cp. Fig.~\ref{fig:j_summary}a). 
Obviously, the choice $f\propto 1/\gamma_h$ seems to be an appropriate choice while $f=const.$ is clearly not supported by the simulations. 
We will still use it as a comparison due the very widespread use of the ponderomotive scaling~\eqref{eqn:ponderomotive}.\\

Finally, we can determine $N_h^{tot}$ using the energy flux conservation Eqn.~\eqref{eqn:Efluxconserve} by identifying $n_h(t)$ with $\partial N_h/\partial t$. 
We can hence readily write
\begin{eqnarray}
N_h^{tot}&=&\frac{\chi a_0^2}{2}\frac{\tau}{\langle f(t)v_h(t)\left(\gamma_h(t)-1\right)\rangle}. 
\label{eqn:Ntot}
\end{eqnarray}\\

We can now evaluate Eqn.~\eqref{eqn:spectrum_general} and give an explicit expression for the spectrum of laser accelerated electrons. 
Applying~\eqref{eqn:gammat}, we obtain the spectrum for modest laser intensities and steep front surface density gradients
\begin{eqnarray}
 \frac {dN_h}{d\gamma_h}&=&N_h^{tot}f  \frac{\gamma_h}{\sqrt{\left(\gamma_h^2-1\right)\left(\hat \gamma^2_h-\gamma_h^2\right)}}
 \label{eqn:spectrum}
\end{eqnarray}
where $\hat \gamma_h\equiv \sqrt{1+\hat a_0^2}$. 

With~\eqref{eqn:gammat_long} the spectrum for ultra-relativistic laser intensities or long preplasma scale lengths is obtained, 
\begin{align}
\nonumber \frac {dN_h}{d\gamma_h}=&N_h^{tot}f\hat a_0
\frac{S^6+8}{2S^5-8S}\sqrt{1-\left(\frac{S^6-8}{6S^3\hat a_0}\right)^2}^{-1}\\
 S=&\sqrt{\gamma_h+1+\sqrt{\gamma_h^2+2\gamma_h-3}}. 
 \label{eqn:spectrum_long}
\end{align}\\

Comparison with the spectra from the simulations yields a reasonable agreement, see Fig.~\ref{fig:spec}. 
We compare the spectra integrated over the ROI for $a_0=10$ (Fig.~\ref{fig:spec}a) and $a_0=30$ (Fig.~\ref{fig:spec}b). 
Especially for $a_0=10$ we see strong modulations at higher energies, for $a_0=30$ the predicted high energy peak is washed out, but otherwise the spectra are described well by Eqn.~\eqref{eqn:spectrum} and~\eqref{eqn:spectrum_long} with  $f\propto \nicefrac{1}{\gamma}$. 
The energy oscillations are a consequence of the interaction of the laser accelerated electrons with the bulk plasma oscillations mentioned before, which is not included in our simple model. 
In fact, just before entering the target the bunch followed almost exactly~\eqref{eqn:spectrum} (Fig.~\ref{fig:spec}c). 

While the interaction with the bulk plasma alters the spectral shape of the $2\omega_0$ bunch, the average energy is not altered significantly. 
This can be seen from Fig.~\ref{fig:spec}d where we show, as a function of the laser strength, the average energies
\begin{equation}
 T_m = \frac{1}{N_h}\int_{m+1}^\infty \gamma_h \frac{dN_h}{d\gamma_h}d\gamma_h-\left(m+1\right)
 \label{eqn:T_m}
\end{equation}
with $N=\int_{m+1}^\infty \nicefrac{dN_h}{d\gamma_h}d\gamma_h$, 
again neglecting the very low energy electrons because they are influenced by dispersion. 
The average electron energies follow almost perfectly  Eqn.~\eqref{eqn:spectrum} for $a_0\ll20$ and  Eqn.~\eqref{eqn:spectrum_long} for $a_0\gg20$, hence again validating the choice $f\propto 1/\gamma$.\\

\begin{figure}
	\includegraphics[width=7cm]{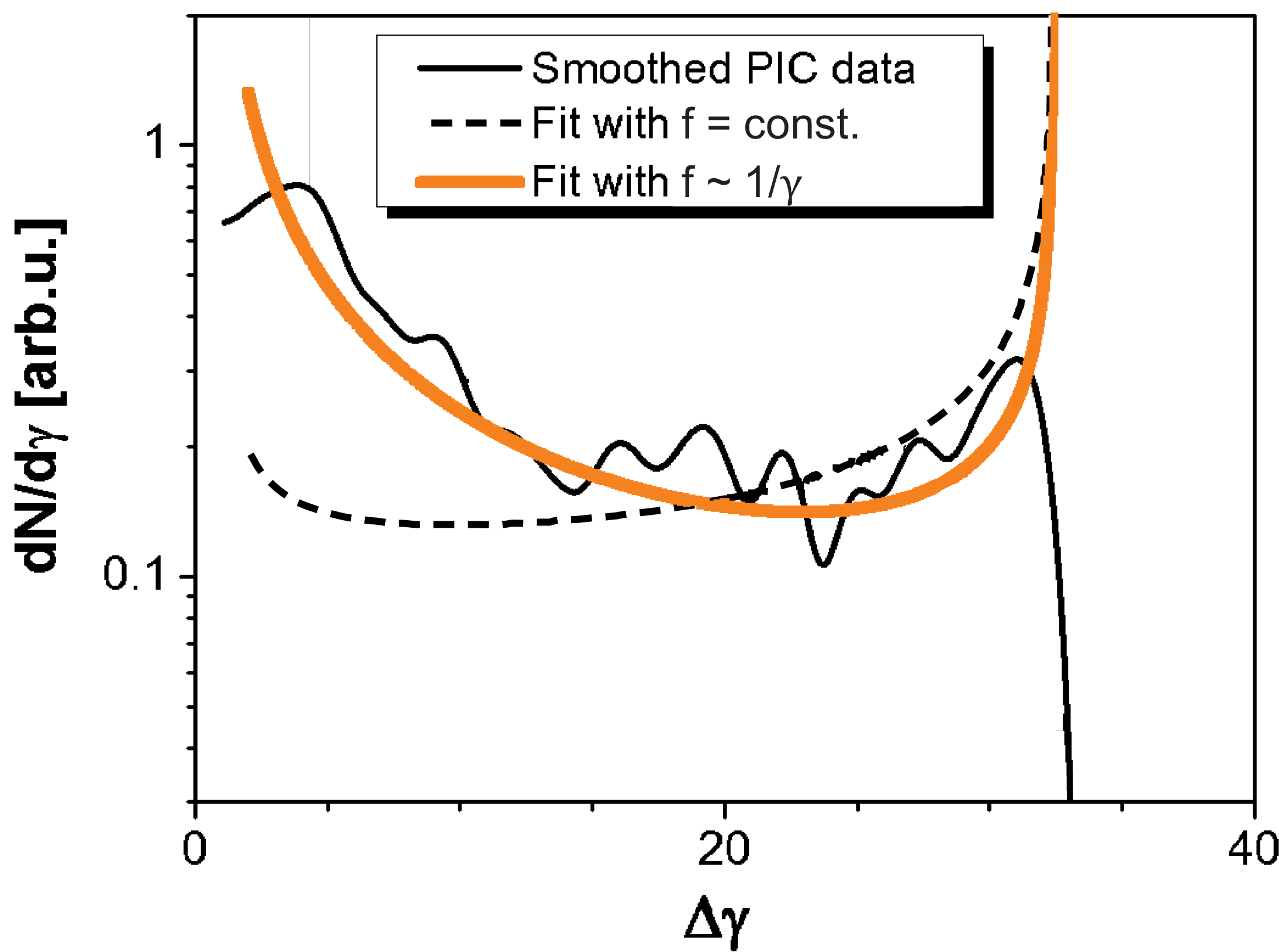}
	\caption{Spectrum of energy $\Delta \gamma$ gained by transverse forces (primarily the $\vec{v} \times \vec{B}$ force) during a quarter laser period for electrons situated in front of the solid density region (neglecting any longitudinal plasma forces, and transverse forces in the bulk and behind the foil) from PIC (black line) and a fit with Eq.~\eqref{eqn:spectrum} and $f\propto 1/{\Delta \gamma}$ (orange line) and $f=const.$ (black dashed line). $a_0=20$ and $\hat{a}=1.6 a_0$.} 
	\label{fig:osci}
\end{figure}

We note, that the simulation setup is chosen deliberately to match the analytical case, i.e. having a plane laser wave interacting with a plane, cold, pre-ionized target. 
If we would have looked at our simulation at a later time, energetic electrons would have been reflected from the foil rear surface and reenter the laser interaction zone. Also, the foil would have been heated and a strong bulk return current would be present. 
All these effects are expected to alter the acceleration scenario and therefore hot electron current generation, e.g. electrons enter the laser field with a non-zero inital velocity. 
Yet, it is interesting to note that a sub-component of the energetic electron current still is accelerated following the same principles as outlined above. 
If we only consider electrons in a bunch that have been accelerated by the laser once (by the $v\times B$ force) and were not part of the bulk return current or have been accelerated and reflected at the target rear before, we can calculate their energy gain $\Delta \gamma=\int_{t_0}^{t_0+\pi/2}a_y(t)v_y(t)dt$ during the quarter laser period of acceleration (where $a_y(t)$ is the transverse electric field at the particle position at time $t$). 
Fig.~\ref{fig:osci} shows the respective spectrum of $\Delta\gamma$ at $t_0=8T$, i.e. 8 laser periods after the laser maximum has reached the target front. 
At that time there exists a strong electro-static surface field which together with strong plasma oscillations, collisions and rear surface fields acts to modify the spectrum from Eqn.~\eqref{eqn:ft}. 
Yet, for the electrons from the surface the spectrum of energy obtained by transverse forces (i.e. the laser) still shows good agreement with the analytical spectrum Eqn.~\eqref{eqn:spectrum}. 
Contrarily, the spectrum and current taking into account all electrons inside a bunch is quite different at such late time and needs a different treatment beyond the scope of this paper. \\

\begin{figure}
	\includegraphics[width=\linewidth]{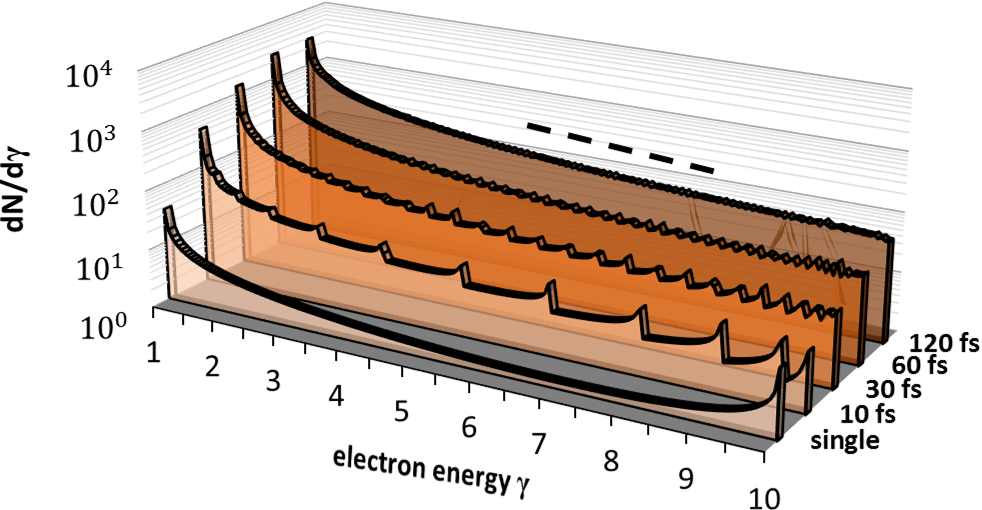}
	\caption{Spectra calculated from Eqn.~\eqref{eqn:spectrum} for $\hat a_0=10$ and Gaussian temporal envelope of $10\unit{fs}$ (light orange), $30\unit{fs}$ (dark orange), $50\unit{fs}$ (brown). The average energy is agrees reasonably with scaling Eqn.~\eqref{eqn:avg_energy_pre} in all cases while the scale length of the tail of the spectrum approaches approximately an exponential form with the scale length given by the ponderomotive potential, Eqn.~\eqref{eqn:ponderomotive}, in the limit of long pulse durations.} 
	\label{fig:exponential}
\end{figure}

It is also worth mentioning that the spectra calculated above are not exponentially decreasing while experimental spectra and simulations in sup-picosecond laser-solid interactions consistently do show exponential behavior in the high energy region\cite{Cowan1999,*Schwoerer2001,*Ewald2002,*Lefebvre2003,*Mishra2009,*Robinson2013a,*Krygier2014,*Chen2015,Chrisman2008}. 
Often, this is ascribed to a thermalization, thus the slope is named temperature and interpreted as the high energy limit of a Maxwellian electron distribution. 
However, two simple arguments show that the distribution of energetic laser accelerated electrons is not thermal: First, the stopping range of MeV electrons in solid matter is on the order of millimeters\cite{star}, and in vacuum behind the target on the path towards the detector thermal equilibration would need much more than the few meters typical distance to the detector -- thus there is simply not enough time for the $\unit{\MeV}$ electrons to equilibrate. 
And secondly, even if the laser accelerated electrons would thermally equilibrate, the distribution should follow a relativistic Maxwellian distribution -- which would not approximate to an exponential curve in the high energy limit.\\
There are now three other explanations that can explain the empiric exponential spectra. 
First, the electron acceleration process itself can show random fluctuations from cycle to cycle, immediately leading to an exponential spectrum\cite{PhysRevLett.44.651} (still not thermal). 
Secondly, the interaction of the laser accelerated electrons with the bulk, i.e. generation of plasma waves, can cause a large amount of randomization of the phase space if the sample is thick enough, from our simulations here we see that this needs a thickness in the order of several micrometers for mildly relativistic short laser pulses. 
Finally, for high contrast lasers and ultra-thin foils where the latter two points do not apply, we now show that one can still find an exponential spectrum with a slope connected to the average electron energy (i.e. similar to a true non-relativistic Maxwellian). \\
If we consider a thin foil (to neglect interaction between laser accelerated electrons with the bulk) and a Gaussian laser pulse temporal envelope, then the spectrum of laser accelerated electrons is dominated by the  controlled direct interaction of the laser with the plasma treated above, with only minor cycle-to-cycle fluctuations. 
In the limit of a slowly varying envelope, each bunch follows the spectral shape derived above, but with different $\hat a_0$ given by the appropriate instantaneous laser envelope. 
The total spectrum then approximately is given by the superposition of all individual bunches. 
The resulting spectrum is shown in Fig.~\ref{fig:exponential} exemplary for $\hat a_0=10$ and various pulse durations. \\
Interestingly, the average energy of relativistic electrons with kinetic energy larger than the rest mass energy is still given in good approximation by Eqn.~\eqref{eqn:avg_energy_pre}, which is true not only for the case of $\hat a_0=10$ shown in the figure, but for all practically relevant values of $\hat a_0$, which therefore can be seen as a good approximation for the relativistic electron energy scaling also for the more realistic Gaussian pulse shapes. \\ 
For sufficiently long pulses the spectrum approaches an exponential in the high energy region. 
Coincidently, in the limit of long Gaussian pulses the \textit{slope} of the high energy tail is approximated with the Wilks ponderomotive scaling Eqn.~\eqref{eqn:ponderomotive}. 
Yet, it is important to realize that this is merely a consequence mainly of our assumption of a Gaussian beam profile and constant laser absorption coefficient throughout the laser interaction. 
It is hence rather a coincidence that the Wilks scaling is recovered; in a realistic situation, e.g. when the laser absorption is a function of many parameters, the high energy spectrum cannot be expected to generally follow an exponential with the scale given by the ponderomotive energy, while our general model can be adapted to take this into account for the average energy. 
%
%

\section{Time averaged laser generated current}
In the following we show how the current density can be modeled for arbitrary $a_0\ll n_e$ including its temporal structure. 
We first compute the average current by explicitly solving the time average integral in Eq.~\eqref{eqn:Efluxconserve}. 
The resulting analytical expressions only depend on the absorption fraction $\chi$ which is the only independent input observable in our model descriptions. 
In a second step we then derive the explicit time dependence of $j_h(t)$. \\

To derive the exact expression we start with rewriting Eqn.~\eqref{eqn:Efluxconserve} and the expression for $\left\langle j_h\right\rangle$, and assume that the electrons inside the plasma move ballisticly in the direction of the laser, $v_z(t)=\nicefrac{\sqrt{\gamma^2(t)-1}}{\gamma_h(t)}$, 
\begin{align}
 \chi \frac{a_0^2}{2}&=\langle n_h\rangle\left\langle f(t) \frac{\left(\gamma_h(t)-1\right)\sqrt{\gamma_h(t)^2-1}}{\gamma_h} \right\rangle
 \label{eqn:new_j_h_both1}\\
 \left\langle j_h\right\rangle&=\langle n_h\rangle\left\langle f(t) \frac{\sqrt{\gamma_h(t)^2-1}}{\gamma_h}\right\rangle.
 \label{eqn:new_j_h_both2}
\end{align}
Eqn.~\eqref{eqn:new_j_h_both1} can be used to eliminate the unknown $\langle n_h \rangle$ in \eqref{eqn:new_j_h_both2}. 
For the mildly relativistic case this can be done analytically. 
Then, $\gamma_h(t)$ is simply the quiver energy \eqref{eqn:gammat}, and with $f(t)\propto 1/\gamma_h(t)$ from \eqref{eqn:ft} one can now easily express the average current as a function of the vacuum field strength $a_0$, and the field strength and energy amplitude at the critical density, $\hat a_0$ and $\hat\gamma_h$
\begin{eqnarray}
\label{eqn:jfinal}
\left\langle j_h\right\rangle & = & \chi \frac{a_0^2}{2}\left(\frac{\tan^{-1}{\left(\hat{a}_0\right)}\hat\gamma_h}{\tanh^{-1}{\left(\nicefrac{\hat{a}_0}{\hat\gamma_h}\right)}}-1\right)^{-1}\\
\nonumber 
& \cong & \chi \left(I_{18}\right)^{0.56}\left(\lambda[\mum]\right)^{-0.88}\cdot 4.5\times 10^{12}\unit{A/cm}^2
\end{eqnarray}
where $I_{18}=I/10^{18}\unit{W/cm}^2$. For comparison, if we would have chosen $f(t)=const.$ the spectrum for Wilks-like scenario would have been obtained, reading
\begin{equation}
\left\langle j_h\right\rangle=\chi \frac{a_0^2}{2}\left(\frac{\hat{a}_0}{\tan^{-1}{\left(\hat{a}_0\right)}}-1\right)^{-1}.
\label{eqn:AvgCurrPond}
\end{equation}
For the ultra-relativistic case no analytical solution exists, but the equations must be solved numerically. \\

\begin{figure}
	\includegraphics[width=8.5cm]{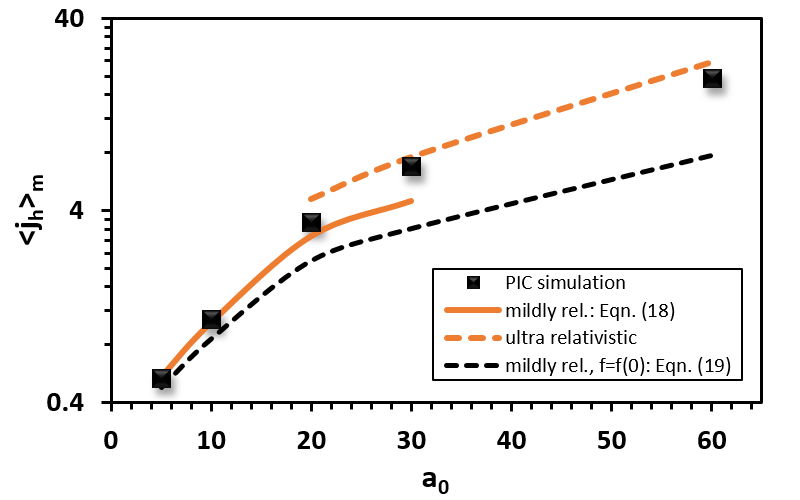}
	\caption{Average current $\left\langle j_h\right\rangle_{m}$ (for $m=1$) of energetic electrons in laser direction inside the ROI. Solid orange line is the numerical model prediction assuming the quiver electron energy scaling \eqref{eqn:gammat} and density evolution \eqref{eqn:ft}, which would correspond to Eqn.~\eqref{eqn:jfinal} for $m=0$. Dashed orange line is the same for the ultra-relativistic case. For comparison. The dashed black line is the current that would have been obtained for \eqref{eqn:EfluxconserveSimpler} in the mildly relativistic case. In all cases we used the laser absorption into energetic electrons extracted from the respective PIC simulation and again assumed $\hat{p_{osc}}=0.9 a_0$. }
	\label{fig:j_hot}
\end{figure}

In order to compare to the PIC simulations, again we compute the current taking only into account the electrons with kinetic energy larger than $m=1$, $\left\langle j_h\right\rangle_{m}$. 
A very good agreement is achieved, again considering the transition between the mildly relativistic and ultra-relativistic laser intensity regimes, see Fig.~\ref{fig:j_hot}. 
The difference in $\langle j_h\rangle$ between the results taking the energy flux density conservation time average correctly, Eqn.~\ref{eqn:Efluxconserve}, versus taking the averages individually for every quantity, Eqn.~\eqref{eqn:EfluxconserveSimpler}, gets increasingly bigger for larger laser strength, as does the difference to the case assuming $f=const.$ (dashed black line). \\

\section{Time dependent laser generated current}
Fig.~\ref{fig:j_summary}c shows the temporal evolution of the current of the simulation with $a_0=10$ by plotting the current as a function of propagation distance into the plasma with the first four half-cycles being convoluted into the period of $0-\pi$. 
Again we only consider energetic electrons with $\gamma-1 > m \equiv 1$, hence $v>0.86 c$. 
Neglecting for now dispersion for the small distance traveled and assuming $v_z\approx c$, space and time are interchangeable and one can think of the figure as temporal evolution of the current of energetic laser generated electrons. \\
Clearly the current is not constant but shows the well-known pronounced $2\omega_0$ structure. 
Moreover, the current shows a similar temporal behavior within a bunch as seen before for the density. 
To calculate the time dependent current $j_h(t)=\langle j_h(t) \rangle f(t)v_z(t)/\langle f(t)v_z(t)\rangle$ we again assume that the electrons move ballistically into the laser direction, $v=v_z$. 
For the mildly relativistic case the current is then calculated to
\begin{equation}
j_h(t)= \langle j_h\rangle \frac{\pi\hat\gamma_h}{\tanh^{-1}\left(\nicefrac{\hat{a}_0}{\hat\gamma_h}\right)}\frac{\left|\hat a_0\sin t\right|}{1+\hat a_0^2\sin^2 t}
\label{eqn:j_final_temp}
\end{equation}
with $\left\langle j_h \right\rangle$ from Eqn.~\eqref{eqn:jfinal} for $\pi\mathrm{k}<t<\pi(\mathrm{k}+0.5)$ where $\mathrm{k}\in \mathbb{N}$ and else $j_h(t)=0$. 
Again, the only free parameter now remains $\chi$, which we extract from the simulation. 
A quantitative comparison of the calculated current with the simulation is shown in Fig.~\ref{fig:j_summary}c. 
Given the simplicity of the model and approximations we made, the agreement with the PIC simulation is remarkable. 

\section{Discussion}

The laser generated hot electron current influences important physical mechanisms, for example the resistive heating of the plasma bulk, magnetic resistive field generation or ion acceleration at the rear. 
The time-dependence of the current has important impact not only due to the modified expression for the average current but also due to the fact that generally current and time enter the models with different non-linearity. 
We want to demonstrate this by computing the resistive heating of the plasma bulk which is dominated by Ohmic heating over drag heating and diffusive heating~\cite{Glinsky1995a}, as well as potentially by anomalous heating~\cite{Sherlock2014,*Kemp2016,*Sherlock2016}. 
For Ohmic heating, the change of bulk electron temperature over time is given by
\begin{equation}
\frac{3}{2}n_eT_e^{3/2}\frac{\delta T_e}{\delta t} = Z \Lambda j_h(t)^2
\label{eqn:ohm}
\end{equation}
with charge state $Z$ and Coulomb logarithm $\Lambda$ and assuming the bulk current to balance the laser generated current~\cite{Glinsky1995a,Bell1997}, which is the case for approximately the magnetic diffusion time~\cite{Davies1997}. 
This was solved in~\cite{sentoku} for constant $j_h$. 
To verify that in the present case heating is dominated by Ohmic heating over the other mechanisms, we want to compare the temperature expected from Ohmic heating to the temperature in the collisional PIC simulation. 
As in the present condition of our PIC simulations $j_h$ is not constant, its full time evolution needs to be considered due to the non-linearity of Eqn.~\eqref{eqn:ohm}. 
Extracting $j_h(t)$ from the simulation with $a_0=10$ for example, we can solve the Ohmic heating equation including the time dependence and compare the calculated temperature with the simulated temperature. 
We find a good agreement with the simulation confirming the validity of the Ohmic heating model for our case and the negligible role of anomalous heating. 

We now want to calculate the bulk heating quantitatively from our model. 
From the discussion before we expect that it is not sufficient to use the average current, or a $2\omega_0$ modulated current without temporal structure within the bunch. 
We therefore have to employ the result of our model from Fig.~\ref{fig:j_summary}. 
The resulting bulk temperature (solid orange line) compares well with the PIC simulation and especially is in much better agreement than using a constant average current during the bunch (orange dashed line). 
For comparison we also show the temperature based on using the simple estimate $j_h=\left\langle \gamma_h\right\rangle n_c=const.$ for $\left\langle \gamma_h\right\rangle$ given by the ponderomotive scaling (dashed black lines) and Eq.~\eqref{eqn:avg_energy_pre} (gray line) which by far underestimate and overestimate the bulk temperature, respectively. 
\begin{figure}
	\includegraphics[width=8.5cm]{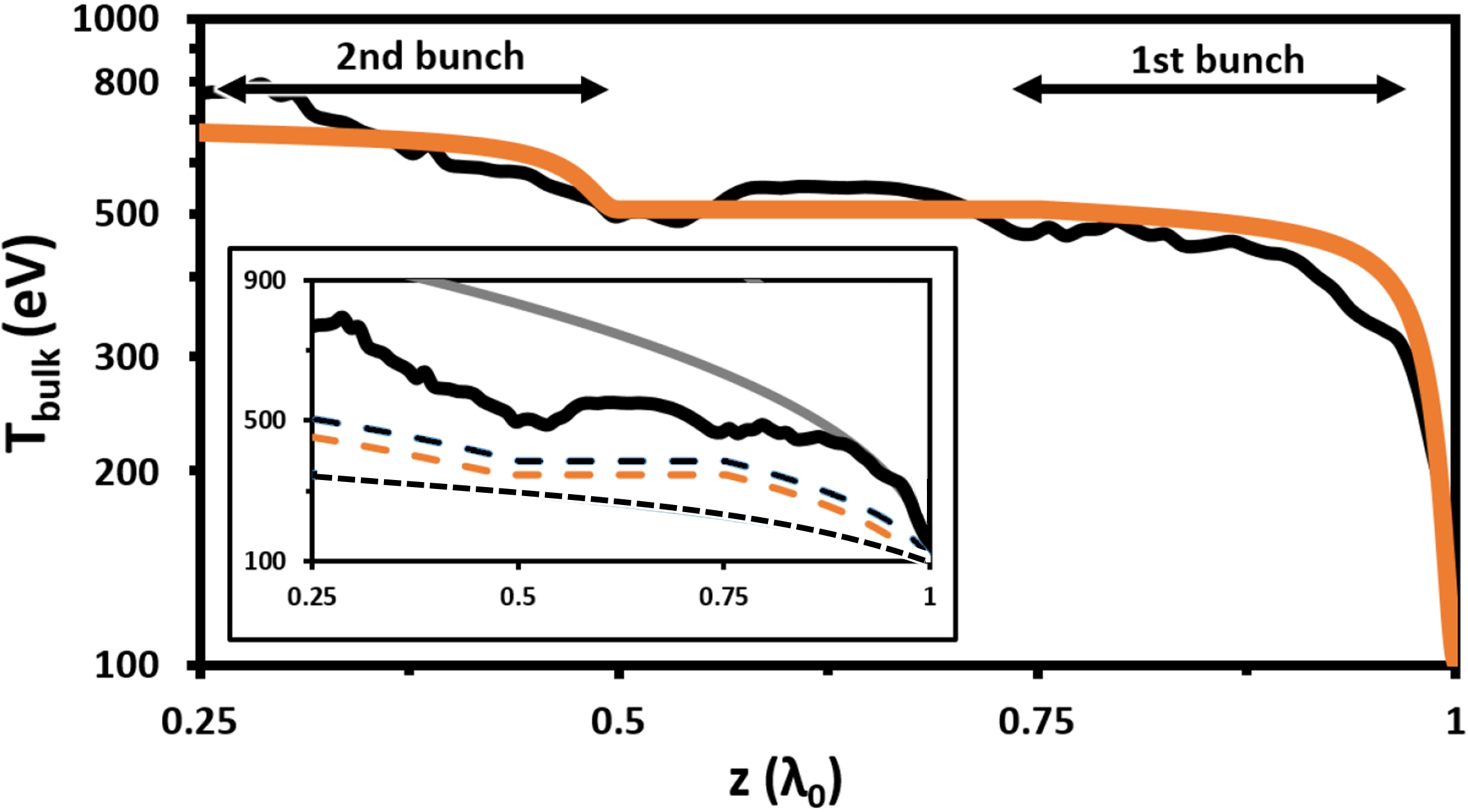}
	\caption{Bulk temperature from a PIC simulation with $a_0=10$ (black solid line) and Ohmic heating~\eqref{eqn:ohm} using the electron current $j_h(t)$ from \eqref{eqn:j_final_temp} (orange solid line). The inset shows for comparison the predictions of only using a constant current during an electron bunch and zero otherwise (with the same average current as above)  (orange dashed); using $j_h=\langle \gamma_h \rangle_m n_c c$ from Eqn.~\eqref{eqn:EfluxconserveSimpler} and ponderomotive energy scaling~\eqref{eqn:ponderomotive}, constant during the bunches and zero otherwise (black dashed line); the same but constant all the time (short-dashed black line); and constant current $j_h=\bar\gamma_h$ with $\bar\gamma_h$ from \eqref{eqn:avg_energy_pre} (gray solid line).}
	\label{fig:ohmic}
\end{figure}

\section{conclusion}
We have demonstrated that the calculation of the laser generated electron current is much more complex than using the simple equation $j_h=\left\langle \gamma_h\right\rangle n_c$. 
First, the spectral distribution had to be taken into account when calculating the average current density using the energy flux conservation. 
Here, in contrast to\cite{Leblanc2014}, we kept the temporal energy evolution as the quiver energy in first approximation with a small correction for ultra-relativistic intensities to account for the longitudinal co-propagation of electrons with the laser fields. i.e. the average electron energy does not depend on the laser absorption in agreement with theory and simulation (e.g.~\cite{wilks1997absorption}). 
This semi-empiric ansatz means that we put all the absorption dependency to the density of laser accelerated electrons, which we describe by Eqn.~\eqref{eqn:ft}. 
Taking the current's temporal dependence is the most dominant effect in Eqn.~\eqref{eqn:Efluxconserve}. 
The average current $\langle j_h\rangle$ is then not given by $\langle \gamma_h \rangle n_c$ or $\bar\gamma_h n_c$ anymore. 
We derived an exact analytical scaling law for the mildly relativistic case, assuming the energy evolution is given by the quiver energy and the electron density is inversly proportional to the energy -- both being in agreement with our simulations. 
For the ultra-relativistic case the same procedure can be applied, but in that case the solution can not be given by a closed analytic expression. 

To derive the temporal evolution of the current one has to take into account the $2\omega_0$ structure imprinted on the current, i.e. the bunch duration of a quarter laser period and separation between the bunches of equal duration. 
The temporal structure within a bunch is of equal importance and directly follows from the temporal evolution of the electron density and velocity. 
For ultra-short laser pulses, only with the full time-dependent modeling non-linear phenomena can be described correctly, which we demonstrated on the example of Ohmic heating. \\

Additionally to our model, effects such as dispersion, front surface fields and interaction with bulk plasma oscillations inside the plasma slab, as well as recirculating laser accelerated electrons can change the spectrum as well as the temporal structure. 
This limits our model to ultra-short time durations, with short propagation distance of electrons in high plasma density and no recirculation, i.e. short laser pulse durations or the early stages of the interaction. 

It is important to point out that our model assumptions and PIC simulations parameters of plane laser waves interacting with a planar cold target and a very short ramp up time of only half a laser period were chosen deliberately to satisfy those strict limitations, as illustrative and easy to study examples of the implications of Eqn.~\eqref{eqn:Efluxconserve} compared to \eqref{eqn:EfluxconserveSimpler}. 

While for many cases our idealizations can be expected to be fulfilled, in a real experiment also the finite laser focal size, longer pulse duration and finite foil thickness might lead to a more complex situation, i.e. $\gamma_h(t)$ may become a more complex multi-parametric function, $f(t)$ may not follow the simple relation given above anymore, and the electron phase space will randomize as e.g. target heating~\cite{Huang2013dynamics,*Huang2016}, instabilities~\cite{sentoku2003high,Macchi2001,Sgattoni2015,Metzkes2015,Kluge2015}, density steepening~\cite{Kemp2008} and refluxing electrons, reflected from the target rear surface~\cite{Mackinnon2002,*Mishra2009,*Kluge2010} have to be taken into account.





\section{Data availability}
The datasets generated and analysed during the current study are available from the corresponding author on reasonable request.

\section{Code availability}
The datasets generated during the current study were produced using the code PICLS, available for review from the corresponding author on reasonable request. 

\begin{acknowledgments}
This work was supported by the German Ministry of Education and Research (BMBF), grant number 03Z1O511. 
\end{acknowledgments}

\bibliography{Mendeley}

\end{document}